\documentclass[11pt]{article}
\usepackage{moriond,epsfig}
\def\Journal#1#2#3#4{{#1} {\bf #2}, #3 (#4)}
\def\PRL{\em Phys. Rev. Lett.}

\def\be{\begin{equation}}
\def\ee{\end{equation}}
\begin{document}
\title{TRANSPORT PHENOMENA IN NANOTUBE QUANTUM DOTS \\ FROM STRONG TO WEAK CONFINEMENT}
\author{\underline{JESPER NYG{\AA}RD}$^1$,
DAVID H.\ COBDEN$^{1,2}$}

\address{$^1$Niels Bohr Institute, {\O}rsted Laboratory, Universitetsparken 5, DK-2100 Copenhagen, Denmark}
\address{$^2$Department of Physics, University of Warwick, Coventry CV4 7AL, UK}

\maketitle\abstracts{
We report low-temperature transport experiments on single-wall nanotubes with metallic leads of varying contact quality, ranging from weak tunneling to almost perfect transmission.  In the weak tunneling regime, where Coulomb blockade dominates, the nanotubes act as one-dimensional quantum dots.  For stronger coupling to the leads the conductance can be strongly enhanced by inelastic cotunneling and the Kondo effect.  For open contacts Coulomb blockade is completely suppressed, and the low-temperature conductance remains generally high, although we often see distinct dips in the conductance versus gate voltage which 
may result from resonant backscattering.
} 

Experiments on single-wall carbon nanotubes have allowed the exploration of numerous quantum transport phenomena for the first time in a molecular system.\cite{springerbook}  Until recently these phenomena were limited to the Coulomb blockade regime owing to the relatively poor (low-transmission) contacts achieved between the nanotubes and the metallic leads.\cite{tans97,nygard99}  However, devices with good (high-transmission, or `open') contacts can now be fabricated~\cite{kong99,liang01} by evaporating metals on top of nanotubes grown by either laser ablation or chemical vapor deposition.  With improved contacts, cotunneling processes and the Kondo effect~\cite{kouwenhoven01} become important, and indeed nanotube devices are a promising system for exploring new extensions of Kondo physics.\cite{nygard00}  Further, once the transmission becomes very high we find behavior similar to that seen very recently by Liang {\em et al.}\cite{liang01} Here {\em dips}\ appear in the conductance which are likely to be caused by resonances in the nanotube which acts as a cavity confined by the weakly reflecting contacts.

Our own devices are based on laser ablation grown single-wall nanotubes, deposited onto a SiO$_2$ substrate after sonication in dichloroethane.  Pure gold leads are evaporated directly onto the tubes using electron-beam lithography and lift-off.  A strongly doped silicon layer underneath the 300 nm thick SiO$_2$ layer is used as a gate electrode. A schematic device is shown in Fig.~1a, where the atomic force microscope (AFM) image reveals a nanotube bundle bridging two Au leads spaced by $L=300$~nm.

\begin{figure}
\begin{center}
\psfig{figure=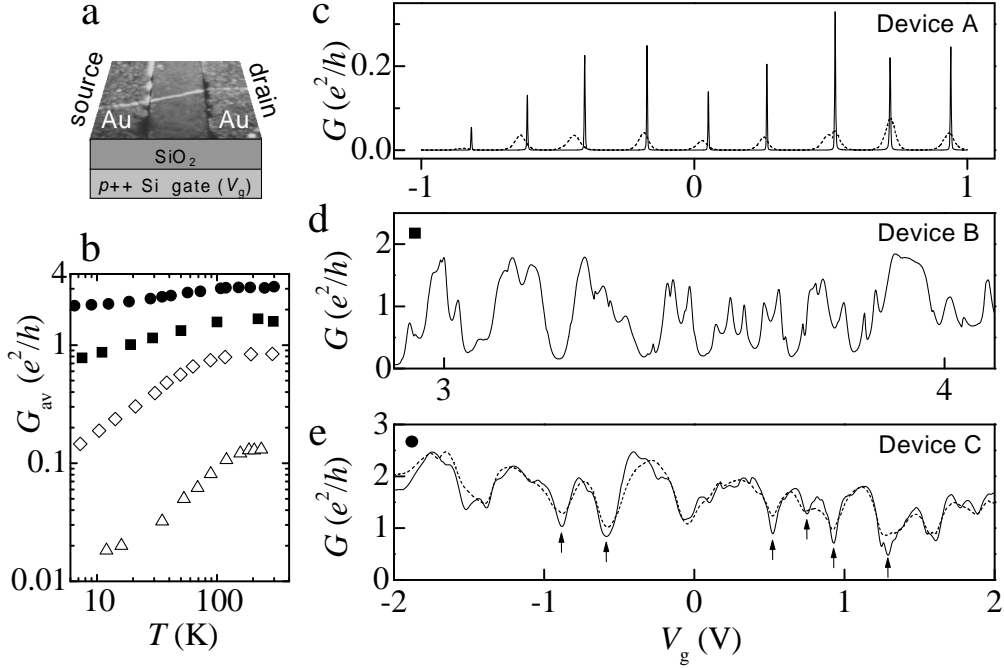,height=9.5cm}
\end{center}
\caption{(a) AFM image of a device. (b) Average conductance $G_{\rm av}$ vs $T$ for devices with different contact transparencies. 
(c)--(e) Conductance $G$ vs gate voltage $V_{\rm g}$.  (c) Device A at 4 K (dotted) and 300 mK (solid).  (d) Device B (squares in b) at 75 mK. (e) Device C (circles in b) at 4.2 K (dotted) and 1.2 K (solid).
\label{fig1}}
\end{figure}

Using this technique the two-terminal conductance at room temperature, $G_{\rm RT}$, is found to vary greatly between devices (for reasons uncertain at present).  We select devices where $G_{\rm RT}$ is independent of gate voltage $V_{\rm g}$, implying that a metallic nanotube with little disorder~\cite{bockrath00} determines the conductance $G$.  In these devices the conductance is limited mainly by the contacts.\cite{bachtold00}  Assuming for simplicity both contacts have transmission probability $P_{\rm C}$, one can show~\cite{datta} that approximately $G_{\rm RT} \leq G_{\rm max} \times P_{\rm C}/(2-P_{\rm C})$, and thus $P_{\rm C} \geq 2/(1+G_{\rm max}/G_{\rm RT})$, where $G_{\rm max}=4e^2/h$ is the maximum possible theoretical conductance of a tube.

In Fig.~1b we plot the average conductance $G_{\rm av}$ vs $T$ for four devices with geometry and size similar to that in Fig.~1a, having $G_{\rm RT}$ ranging from $0.015~e^2/h$ ($P_{\rm C}<0.01 $) to $3.1~e^2/h$ ($P_{\rm C}\sim 0.9$). The two devices with $ G_{\rm RT}< e^2/h$ (open symbols) show the strongest variation, their conductance decreasing roughly as a power law, $G\sim T^{0.7}$, below 100 K.  This power law behavior, generic in low-conductance nanotube devices, can be consistently explained by the suppression of tunneling between the leads and the interacting one-dimensional electron system in the nanotube (thought to be a Luttinger liquid) at low energies.\cite{bockrath99}  The other two devices with higher $G_{\rm RT}$ (solid symbols) have much weaker $T$ dependences and do not show such power laws.

At low $T$ all the devices exhibit rapid, reproducible variations in the conductance $G$ as a function of gate voltage $V_{\rm g}$.  Figs.~1c--e show the behavior of three devices spanning a wide range of contact transmission.
For the low-conductance device labeled A (Fig.~1c), with $G_{\rm RT}=0.3~e^2/h$ ($P_{\rm C}\sim 0.15$), Coulomb blockade~\cite{tans97} (CB) sets in thoroughly, producing a series of roughly periodic, isolated CB peaks.  This is the norm for low-conductance devices.  For device B (Fig.~1d), which has $G_{\rm RT}=1.6~e^2/h$ ($P_{\rm C}\sim 0.6$), the structure is still peak-like, but the peaks are less regular and $G$ never vanishes between them even at 75 mK.  For the highest conductance device, C, with $G_{\rm RT}=3.1~e^2/h$ ($P_{\rm C}\sim 0.9$), the fluctuations are weaker and no longer peak-like (Fig.~1e).  In fact, in this case we can identify a series of well defined {\em dips} in $G$ vs $V_{\rm g}$ which deepen as $T$ is decreased from 4.2~K to 1.2~K, such as those indicated by arrows in Fig.~1e.

\begin{figure}
\begin{center}
\psfig{figure=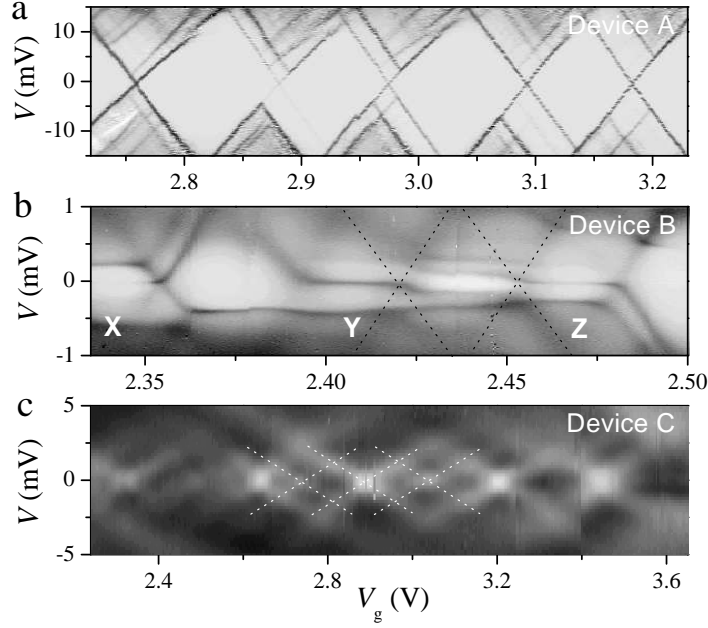,height=8.5cm}
\end{center}
\caption{(a) Differential conductance ${\rm d}I/{\rm d}V$ vs $V_{\rm g}$ and bias $V$ for the same devices as in Fig.~1c--e. Measurements at 75 mK (a), 75 mK (b), and 1.2 K (c). Darker means higher conductance. 
\label{fig2}}
\end{figure}

Nonlinear transport measurements reveal the key differences between these devices with different contact transmissions.  In Fig.~2 the differential conductance ${\rm d}I/{\rm d}V$ is plotted in greyscale vs source-drain bias $V$ and $V_{\rm g}$.  For device A (Fig.~2a) the result is a standard Coulomb blockade diamond pattern, with zero conductance inside the diamonds.\cite{tans97,nygard99}  The gate coupling $C_{\rm g}/C \sim 0.13$ is typical for devices of this size  ($C_{\rm g}$ is the capacitance to the gate and $C$ the total capacitance).  On the other hand, for both devices B and C, $C_{\rm g}/C\leq 0.025$, implying that the potential in the nanotube is much less sensitive to the gate voltage when the contacts are more open, in spite of the similar geometry to device A.  The mean level spacing in device A of $\Delta E \sim 2$~meV is about one sixth of the charging energy $U$ and is consistent with the estimate~\cite{tans97} $\Delta E\approx \hbar v_{\rm F}\pi/L$, where the separation $L$ of the leads defines the effective length of the dot. This implies that the localization length in the tube is longer than $L$.

For device B (Fig.~2b), Coulomb diamonds can again be discerned, as indicated by the dotted lines, showing that CB is still effective enough to make the electron number $N$ on the nanotube a well defined quantity.  The diamonds are however much less distinct than in device A.  This is explained by the increased level broadening and cotunneling resulting from the stronger lead coupling.  What stands out most though is the presence of horizontal features  of enhanced ${\rm d}I/{\rm d}V$ at a certain $V$.  In particular, there are ridges at $V=0$ in the diamonds marked X (faint), Y, and Z.  In Fig.~3a we show for the same $V_{\rm g}$ range the small-bias conductance $G$ at a series of temperatures.  At the highest $T$ (dotted line) there are fairly regular 
CB oscillations.  However, whereas for regular CB $G$ decreases exponentially in the valleys as $T$ is lowered, this is not the case in valleys X, Y and Z.  In fact, for valley Y, which shows the strongest zero-bias ridge in Fig.~2b, $G$ {\em increases} logarithmically with decreasing $T$ as shown in Fig.~3b.

\begin{figure}
\begin{center}
\psfig{figure=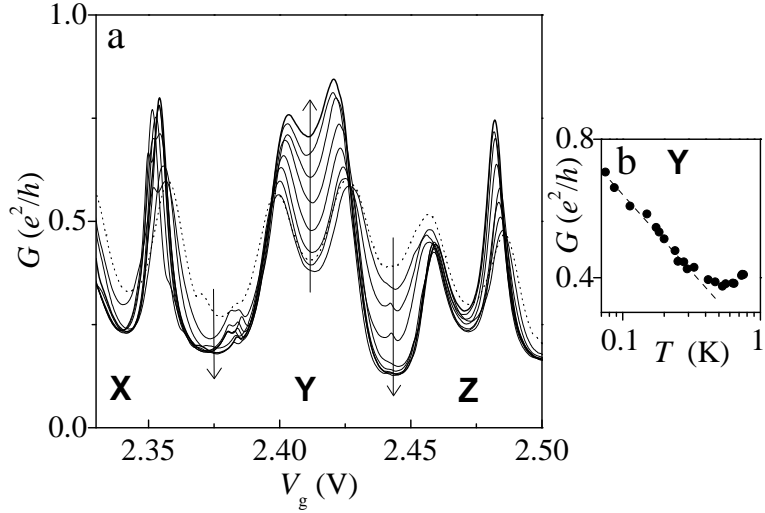,height=7cm}
\end{center}
\caption{(a) $G$ vs $V_{\rm g}$ for the device B (Fig.~2b), here at 75 (thick line), 85, 115, 175, 240, 295, 420, 590, and 730 mK (dotted). 
(b) $G$ vs $T$ for valley Y in a. The dotted line indicates $\log T$ behavior.
\label{fig3}}
\end{figure}

The zero-bias features and their anomalous $T$ dependence are signatures of the Kondo effect in a quantum dot, which becomes significant along with cotunneling as the coupling to the leads increases.\cite{kouwenhoven01,nygard00}  When the electron number $N$ is odd (valleys X,Y and Z), the total spin of the nanotube dot must be at least 1/2.  In this case cotunneling processes can occur which flip the spin of the dot.  These allow electrons in the two leads to become correlated, and a Kondo resonance appears at the Fermi energy resulting in enhanced transmission rates between the contacts when $T$ is below the Kondo temperature $T_K$.  On the other hand, in the even-$N$ valleys, the ground state spin is usually zero~\cite{cobden98} and no Kondo effect is seen at zero bias.

Finally, for device C (Fig.~2c) no Coulomb diamonds can be seen, as expected because the contacts are so transparent ($P_{\rm C}\sim 0.9$) that there should be no Coulomb blockade.  However, we notice that there do still exist crosses in this greyscale plot, as indicated by dotted lines, but that they are {\em white}, and are centered on the {\em dips} in $G$ rather than the peaks. In the open regime, such patterns can be generated by Fano resonances, as reported in semiconductor quantum dots by G{\"o}res {\em et al.}\cite{fano}  Such resonances must be formed from the interference between two transmitting modes, which in the case of nanotubes could be the $K$ and $K^\prime$ points in the lowest 1D subband.  The spacing of the dips in energy deduced from this plot is $\sim$1--2~meV, which is roughly the same as the level spacing in device A.  This is again consistent with the absence of CB, ie negligible charging effects, assuming these features are still related to electron states confined, however weakly, within the tube.

Very recently Liang {\em et al.}\ have seen similar, highly regular `inverted diamond' patterns using devices with $G_{\rm RT}\sim 1.5$--$2.5~e^2/h$ based on nanotubes grown in-situ by chemical vapor deposition.\cite{liang01}  They explain the patterns as resulting from a clean metallic nanotube acting as an etalon-like resonant cavity with the contacts as weakly reflecting mirrors which scatter between the $K$ and $K^\prime$ modes.  They find that the spacing of the dips varies with $1/L$, adding further weight to the idea that open metallic nanotubes are ballistic phase-coherent one-dimensional wires.

In summary, 1D transport can now be studied in single-wall nanotubes throughout the range from closed 1D quantum dots through the cotunneling and Kondo regime to the limit of open 1D channels with weak backscattering.

\section*{Acknowledgments}
We thank M.~Bockrath, H.~Bruus, K.~Flensberg, P.E.~Lindelof for discussions, and R.E.~Smalley and A.~Rinzler for supplying the nanotube material.  The work was carried out in the NBI Nanolab and supported the Danish Technical Research Council.


\end{document}